\documentclass[aps,prl,twocolumn,superscriptaddress,showpacs,floatfix,longbibliography]{revtex4-2}

\makeatletter
\def\@bibdataout@aps{%
\immediate\write\@bibdataout{%
@CONTROL{%
apsrev41Control%
\longbibliography@sw{%
    ,author="08",editor="1",pages="1",title="0",year="1"%
    }{%
    ,author="08",editor="1",pages="1",title="",year="1"%
    }%
  }%
}%
\if@filesw \immediate \write \@auxout {\string \citation {apsrev41Control}}\fi
}
\makeatother

\usepackage{graphicx}
\usepackage{dcolumn}
\usepackage{bm}

\usepackage{multirow}

\usepackage{amssymb}
\usepackage{amsmath}
\usepackage{graphicx}
\usepackage{xcolor}
\usepackage[colorlinks,citecolor=blue,linkcolor=red,urlcolor=blue]{hyperref}

\begin{document}
\title{Driven Critical Dynamics in Measurement-induced Phase Transitions}
\author{Wantao Wang}
\thanks{These two authors contributed equally to this work.}
\affiliation{Guangdong Provincial Key Laboratory of Magnetoelectric Physics and Devices, Sun Yat-Sen University, Guangzhou 510275, China}
\affiliation{School of Physics, Sun Yat-Sen University, Guangzhou 510275, China}
\author{Shuo Liu}
\thanks{These two authors contributed equally to this work.}
\affiliation{Institute for Advanced Study, Tsinghua University, Beijing 100084, China}
\author{Jiaqiang Li}
\affiliation{Guangdong Provincial Key Laboratory of Magnetoelectric Physics and Devices, Sun Yat-Sen University, Guangzhou 510275, China}
\affiliation{School of Physics, Sun Yat-Sen University, Guangzhou 510275, China}
\author{Shi-Xin Zhang}
\affiliation{Institute of Physics, Chinese Academy of Sciences, Beijing 100190, China}
\author{Shuai Yin}
\email{yinsh6@mail.sysu.edu.cn}
\affiliation{Guangdong Provincial Key Laboratory of Magnetoelectric Physics and Devices, Sun Yat-Sen University, Guangzhou 510275, China}
\affiliation{School of Physics, Sun Yat-Sen University, Guangzhou 510275, China}

\date{\today}

\begin{abstract}
Measurement-induced phase transitions (MIPT), characterizing abrupt changes in entanglement properties in quantum many-body systems subjected to unitary evolution with interspersed projective measurements, have garnered increasing interest. In this work, we generalize the Kibble-Zurek (KZ) driven critical dynamics that has achieved great success in traditional quantum and classical phase transitions to MIPT. By linearly changing the measurement probability $p$ to cross the critical point $p_c$ with driving velocity $R$, we identify the dynamic scaling relation of the entanglement entropy $S$ versus $R$ at $p_c$. For decreasing $p$ from the area-law phase, $S$ satisfies $S\propto \ln R$; while for increasing $p$ from the volume-law phase, $S$ satisfies $S\propto R^{1/r}$ in which $r=z+1/\nu$ with $z$ and $\nu$ being the dynamic and correlation length exponents, respectively. Moreover, we find that the driven dynamics from the volume-law phase violates the adiabatic-impulse scenario of the KZ mechanism. In spite of this, a unified finite-time scaling (FTS) form can be developed to describe these scaling behaviors. Besides, the dynamic scaling of the entanglement entropy of an auxiliary qubit $S_Q$ is also investigated to further confirm the universality of the FTS form. By successfully establishing the driven dynamic scaling theory of this newfashioned entanglement transition, we bring a new fundamental perspective into MIPT that can be detected in fast-developing quantum computers.
\end{abstract}

\maketitle

{\bf Introduction}---Measurement is not only a fundamental issue in quantum theory, but also an indispensable element in the development of quantum technology. Recently, the interplay between measurements and unitary evolution in quantum circuits has become a very active subject of research. On the one hand, unitary evolution increases the entanglement of generic isolated quantum many-body systems to a volume-law scaling by scrambling quantum information throughout the whole system~\cite{Calabrese_2005,HusePhysRevLett111127205,NahumPhysRevX7031016,NahumPhysRevX8021014,SondhiPhysRevX8021013,TurkeshiPhysRevLett131230403}. On the other hand, measurement is an intrinsically nonunitary operation due to the wave-function collapse in the process. Local measurements generally decrease entanglement since the subsystem under measurement can be disentangled from the rest~\cite{Daviesbook,Zurekbook}. In monitored quantum circuits with unitary evolution interspersed by measurements, the competition between them leads to the measurement-induced phase transition (MIPT), separating a volume-law phase of entanglement entropy with infrequent measurements to an area-law phase with frequent measurements~\cite{LiydPhysRevB98205136,NahumPhysRevX9031009,LiydPhysRevB100134306}. Such MIPT is believed to be ubiquitous in fast-developing newfashioned quantum devices and its theory has profound connections to percolation and conformal field
theory, as well as threshold theorems in fault-tolerant quantum computing~\cite{LiydPhysRevB98205136,NahumPhysRevX9031009,LiydPhysRevB100134306,ChanPhysRevB99224307,VasseurPhysRevB100134203,HusePhysRevX10041020,HusePhysRevLett125070606,HusePhysRevB101060301,ChoiPhysRevLett125030505,JiancmPhysRevB101104302,BaoPhysRevB101104301,LiusPhysRevB107L201113,IppolitiPhysRevX11011030,TurkeshiPhysRevB106214316,TurkeshiPhysRevB102014315,LiydPhysRevB109174307,LiusPhysRevB110064323,LiusPhysRevLett132240402,CecilePhysRevResearch6033220,XuckPRXQuantum4030317,AshidaPhysRevB110094404,AliceaPhysRevX13041042,MoghaddamPhysRevLett131020401,XuckPhysRevB109035146,FisherPhysRevLett130220404,PolmannPRXQuantum5010309,Lavasaninpexp,Noelnpexp,Koh2023,Hoke2023,PhysRevLett.129.080501, PhysRevResearch.3.023200,kelly2024generalizingmeasurementinducedphasetransitions,qian2024coherentinformationphasetransition}.

Universal nonequilibrium dynamics near a critical point is one of the central issues in traditional phase transitions~\cite{Hohenberg1977rmp,Rigol2016review,Mitra2018arcmp,delcamporev,Polkovnikov2011rmp,Dziarmaga2010review}. For the driven dynamics by linearly changing the distance to the critical point $g$ with a velocity $R$ to cross the critical point, the celebrated Kibble-Zurek (KZ) mechanism provides a unified adiabatic-impulse scenario to understand the scaling behaviors after the quench by comparing the time distance, $g/R$, with the reaction time $\zeta\propto \xi^z$, in which $\xi\propto |g|^{-\nu}$ being the correlation length and $\nu$ and $z$ being the correlation length exponent and dynamic exponent, respectively~\cite{Kibble1976,Zurek1985}. A crucial prerequisite in this scenario is the existence of an initial adiabatic stage, wherein $g/R>\zeta$ and the system evolves along the equilibrium state, since its border with the intrinsically nonequilibrium impulse region, wherein $g/R<\zeta$, gives rise to a frozen time $\hat{t}$, which imposes a spatial separation of phase domains, yielding the KZ scaling of the topological defects~\cite{Kibble1976,Zurek1985,delcamporev,Polkovnikov2011rmp,Dziarmaga2010review}. Furthermore, a finite-time scaling (FTS) theory was proposed to generalize the original KZ scaling to the whole driven process for more general quantities and driving styles~\cite{Zhifangxu2005prb,Gong2010njp}. Both the KZ mechanism and the FTS have aroused intensive investigations from both theoretical and experimental aspects, exerting far-reaching significance in both classical and quantum phase transitions~\cite{Gong2010njp,Kibble1976,Zurek1985,Zoller2005prl,Yin2014prb,Zhifangxu2005prb,huangyy2014prb,Dziarmaga2005prl,Du2023,Ko2019,PhysRevLett.129.227001,sciadv.aba7292,science.abo6587,science.abq6753,PRXQuantum,Sandvik2015prl,Clark2016science,Feng2016prb,Yin2017prl,Keesling2019,Ebadi2021,king2023nature,GarciaPhysRevB110125113,PhysRevB.106.L041109}, and recently, even extending their fabulous power to state preparations and detections of critical properties on rapidly developing quantum computers~\cite{king2023nature,Keesling2019,Ebadi2021,GarciaPhysRevB110125113,PhysRevB.106.L041109}.

As an intrinsic nonequilibrium phase transition, the MIPT occurs in the individual quantum trajectory but cannot be detected in the mixed state averaged over measurement outcomes~\cite{LiydPhysRevB98205136,NahumPhysRevX9031009,LiydPhysRevB100134306}. Accordingly, the MIPT apparently goes beyond the conventional paradigms of both classical phase transitions based on the usual statistical ensemble and quantum phase transitions on the ground state. Given the profound meaning and important application of the KZ dynamics, it is natural to ask whether the driven dynamic scaling behaviors also emerge in the MIPT. If so, how to characterize it? In particular, are the original KZ mechanism and the FTS still applicable?

To answer these questions, we here investigate the driven critical dynamics in MIPT~\cite{LiydPhysRevB98205136,LiydPhysRevB100134306}. By linearly changing the measurement probability $p$ with different driving velocities $R$ to cross the transition point $p_c$ in a $(1+1)$D hybrid stabilizer circuit, we discover rich scaling behaviors for different initial states. For decreasing $p$ from the area-law phase, we find that the dynamics obeys the adiabatic-impulse scenario of the KZ mechanism~\cite{Kibble1976,Zurek1985,delcamporev,Polkovnikov2011rmp,Dziarmaga2010review} and $S$ for a subsystem $A$ satisfies $S\propto \ln R$ at $p_c$ and is almost independent of the size of $A$, denoted as $|A|$, for large $R$ and large $|A|$. In contrast, for increasing $p$ from the volume-law phase, we find that the adiabatic-impulse scenario breaks down since in the initial stage the system evolves non-adiabatically along an $R$-dependent quasi-steady state rather than the steady state without external driving. Moreover, at $p_c$, we observe that $S$ satisfies $S\propto R^{1/r}$ ($r=z+1/\nu$) for large $R$ and strongly depends on $|A|$. Despite these differences, we find that the dynamics of $S$ can be well described by a unified FTS form featured by $R$ as a relevant scaling variable. We also show that the FTS form can be generalized to other quantities. Through successfully establishing the driven dynamic scaling theory of the newfashioned entanglement transition, we not only generalize the conventional theory of driven dynamics but also bring a new fundamental perspective into MIPT that could be detected in fast-developing quantum simulators~\cite{Noelnpexp,Koh2023,Hoke2023}.


\begin{figure}[tbp]
\centering
  \includegraphics[width=\linewidth,clip]{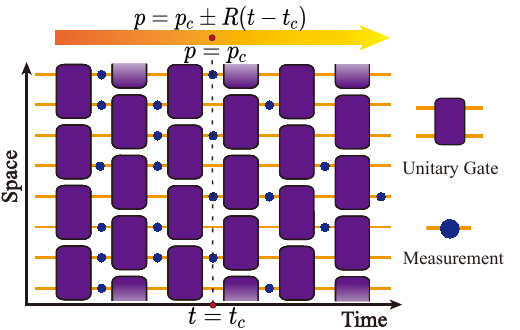}
  \vskip-3mm
  \caption{{\bf Sketch of driven critical dynamics in MIPT.} Purple rectangles represent two-site
  random Clifford gates and blue circles denote local projective measurements that are performed with a probability $p$. The unit of time is chosen as two evolution steps. Each step comprises a measurement layer and a unitary layer. Starting from the steady state of $p_{0}$ which is far from $p_c$, $p$ is changed versus the time $t$ as $p=p_c\pm R(t-t_c)$ to drive the system to cross its critical point at $p_c$ with the corresponding time $t_c$.
  }
  \label{fig:quench}
\end{figure}

{\bf Model and steady-state properties}--- We focus on the $(1+1)$D circuits with the regular brick-wall structure. The time unit is chosen as two evolution steps with each step comprising (i) a unitary layer of random Clifford gates and (ii) a measurement layer stochastically projecting the system qubits~\cite{LiydPhysRevB98205136,LiydPhysRevB100134306}. In the measurement layer,
the projective measurement occurs on each qubit with probability $p$ and then the projector operator $P^{\pm}\equiv (1 \pm Z)/2$ is randomly applied to the spatial wave function according to the Born rules, followed by a normalization procedure~\cite{LiydPhysRevB98205136,LiydPhysRevB100134306}. 
In the unitary layer, the unitary operators in each step are given by $U(t)=\left[\prod_{x-{\rm odd}} U_{(x,x+1),2t}\right]\left[\prod_{x-{\rm even}} U_{(x,x+1),2t+1}\right]$ starting from $t=0$. The periodic boundary condition is imposed in the spatial direction.

It was shown that after an equilibration time scale, the system can enter different steady states depending on  $p$. A critical measurement probability $p_c=0.159 95(10)$ separates a volume-law phase for small $p$ and an area-law phase for large $p$~\cite{TurkeshiPhysRevB106214316}. Near this MIPT, the entanglement entropy $S$ for a subsystem $A$ in a large system obeys a scaling form $S(p,A)=\alpha {\rm ln}|A|+F(g|A|^{1/\nu})$~\cite{LiydPhysRevB100134306}, in which $g=p-p_c$, $\alpha=1.57(1)$~\cite{TurkeshiPhysRevB106214316}, $\nu=1.260(15)$~\cite{TurkeshiPhysRevB106214316}, and $F$ is a scaling function. At the critical point $p_c$, $S\propto {\rm ln|A|}$, while $S\propto {\rm ln}\xi \propto {\rm ln}(g^{-\nu})$ in the area-law phase and $S\propto |A|$ in the volume-law phase~\cite{LiydPhysRevB98205136,LiydPhysRevB100134306}.


{\bf General dynamic scaling form}--- Now we consider the driven critical dynamics of the MIPT for the above circuit model under a linear change in the measurement probability $p$ as $p=p_c\pm R(t-t_c)$ with $t_c$ being the time at $p_c$ and with an initial $p_0$ which is far away from $p_c$. According to FTS, near the critical point, $R$ enters the scaling form as an indispensable scaling variable. By imposing a scale transformation on $g$ with a rescaling factor $b$, one finds that $gb^{1/\nu}=Rb^r(t-t_c)b^{-z}$ with $r=z+1/\nu$ ($z=1$ for present case) being the dimension of $R$~\cite{Zhifangxu2005prb,Gong2010njp}. Accordingly, incorporating the steady-state scaling form of $S$~\cite{LiydPhysRevB100134306}, we obtain the FTS form of $S$ for $L\rightarrow \infty$ as
\begin{equation}
S(R,g,|A|)=\alpha {\rm ln}|A|+G(R|A|^{r},g|A|^{1/\nu}).
\label{genscaling}
\end{equation} 

Some remarks on Eq.~(\ref{genscaling}) are discussed as follows. (i) For finite $L$, the size effects can be readily taken into account with $L$ included (see supplemental material (SM)). (ii) For different initial states, the scaling function $G$ can be different. In the following, we will see that different asymptotic behaviors of $G$ appear for area-law and volume-law initial states. (iii) Eq.~(\ref{genscaling}) can be generalized for other quantities, such as the entanglement entropy of an auxiliary qubit, $S_Q$~\cite{HusePhysRevX10041020,HusePhysRevLett125070606}, and the tripartite quantum mutual information $I_3$~\cite{HusePhysRevB101060301} (see SM). (iv) In usual phase transitions~\cite{Kibble1976,Zurek1985,delcamporev,Polkovnikov2011rmp,Dziarmaga2010review}, when $g>R^{1/\nu r}$, i.e., $t-t_c>\hat{t}$, the system evolves along the equilibrium state in the adiabatic stage and the FTS form restores the equilibrium limit. In sharp contrast, in the following, we will see that this is not the case for the MIPT when the driving starts from the volume-law phase, although Eq.~(\ref{genscaling}) is always holding.

\begin{figure}[tbp]
\centering
  \includegraphics[width=\linewidth,clip]{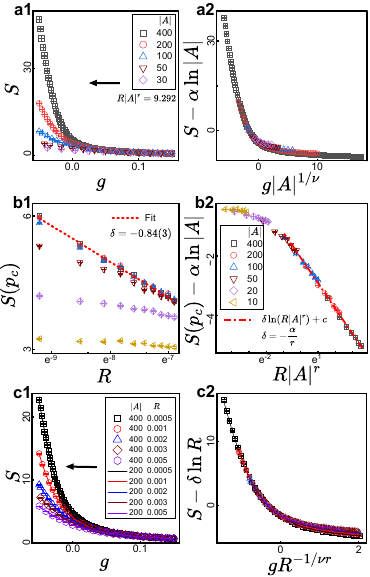}
  \vskip-3mm
  \caption{{\bf Driven critical dynamics of entanglement entropy $S$ from the area-law phase.} (a) For a fixed $R|A|^r=9.292$, curves of $S$ for different subsystem $A$ versus $g$ before (a1) and after (a2) rescaling. (b) Curves of $S$ versus $R$ at $p_c$ for different $A$ before (b1) and after (b2) rescaling. Semi-logarithmic scales are used in (b). Logarithmic fitting (short dash) in (b1) gives $\delta=-0.84(3)$, which is close to $\delta=-0.87(1)$ with $\nu$ and $\alpha$ set as input. Curve of $(\delta {\rm ln}(R|A|^r)+c)$ (dash dot) is plotted for comparison in large $R|A|^r$ region. (c) Curves of $S$ versus $g$ for different $R$ with two fixed $|A|$ before (c1) and after (c2) rescaling. The arrows indicate the driving direction.
  }
  \label{fig:quench1}
\end{figure}

{\bf Driven dynamics from area-law phase}--- Let us begin with the driven dynamics from an initial area-law steady state. The measurement probability is changed according to $p=p_c- R(t-t_c)$ with $p_0=0.30995$. The lattice size is fixed as $L=1024$, for which we have checked that the size effects can be neglected. At first, we examine Eq.~(\ref{genscaling}). Fig.~\ref{fig:quench1} (a) displays the evolution of $S$ for an arbitrary fixed $R|A|^r$. We find in Fig.~\ref{fig:quench1} (a2) that curves of $(S-\alpha{\rm ln}|A|)$ versus $g|A|^{1/\nu}$ collapse well onto each other, confirming that the dynamics of $S$ can be described by Eq.~(\ref{genscaling}).

Moreover, for large $g$, we find that $S$ is almost independent of $R$ (or $|A|$ for fixed $R|A|^r$) and evolves adiabatically along the steady state. As $g$ becomes smaller, curves of $S$ begin to depart from each other. These results demonstrate that the driven dynamics of MIPT from the area-law phase conforms to the adiabatic-impulse scenario of the KZ mechanism~\cite{Kibble1976,Zurek1985,delcamporev,Polkovnikov2011rmp,Dziarmaga2010review}.

Then we explore the scaling property of $S$ at $p_c$. In this case, Eq.~(\ref{genscaling}) reduces to $S(p_c)=\alpha {\rm ln}|A|+G_0(R|A|^r)$. Fig.~\ref{fig:quench1} (b1) shows that for large $|A|$ and large $R$, $S\propto {\rm ln}R$ and scarcely depends on $|A|$. This requires $G_0$ to obey $G_0(R|A|^r)= \delta {\rm ln}(R|A|^r)+c$ ($c$ is a nonuniversal constant) with $\delta=-\alpha/r$ such that the $|A|$-dependent term can be cancelled. In this way, we obtain the scaling relation of $S(p_c)$ for large $R$,
\begin{equation}
S(p_c,R)= \delta {\rm ln}R+c.
\label{arealargeR}
\end{equation}
Eq.~(\ref{arealargeR}) can be verified by the consistency between the $\delta$ value with $\nu$ and $\alpha$ set as input and the coefficient of $S$ versus ${\rm ln}R$ obtained from logarithmic fitting for large $|A|$ as shown in Fig.~\ref{fig:quench1} (b1). Moreover, the scaling collapsing curve of $(S(p_c)-\alpha {\rm ln}|A|)$ versus $R|A|^r$, shown in Fig.~\ref{fig:quench1} (b2), just corresponds to $G_0(R|A|^r)$. From Fig.~\ref{fig:quench1} (b2) one finds that for large $R|A|^r$, $G_0(R|A|^r)$ matches with $\delta {\rm ln}(R|A|^r)+c$; while for small $R|A|^r$, $G_0(R|A|^r)$ tends to saturate to the equilibrium scaling. These results confirm Eq.~(\ref{arealargeR}) and the above scaling analyses.

Moreover, we show that the scaling properties dominated by $R$ also exist in the driven process. Fig.~\ref{fig:quench1} (c1) shows that for large $R$ and large $|A|$ the driven dynamics of $S$ for different subsystem $A$ is almost independent of $|A|$.  Accordingly, combining Eq.~(\ref{arealargeR}) with Eq.~(\ref{genscaling}), we obtain the FTS form of $S$ as
\begin{equation}
S(p,R)= \delta {\rm ln}R+H(gR^{-1/\nu r}),
\label{arealargeR1}
\end{equation}
in which $|A|$ disappears. This $|A|$-independent scaling form is verified in Fig.~\ref{fig:quench1} (c2), which shows that the rescaled curves of $(S-\delta {\rm ln}R)$ versus $gR^{-1/\nu r}$ collapse well onto each other for different $|A|$. Note that the dynamic scaling region governed by Eq.~(\ref{arealargeR1}) can even extend into the volume-law phase for $p<p_c$ as shown in Fig.~\ref{fig:quench1} (c2).

To understand Eqs.~(\ref{arealargeR}) and~(\ref{arealargeR1}), one notices that for driving from the area-law side, scaling properties in the critical region are controlled by the frozen time $\hat{t}$ as a result of the critical slowing down~\cite{Kibble1976,Zurek1985,delcamporev,Polkovnikov2011rmp,Dziarmaga2010review}. The correlation length at $\hat{t}$ obeys $\xi(\hat{t})\propto |\hat{g}|^{-\nu}\propto R^{-1/ r}$~\cite{Zhifangxu2005prb,Gong2010njp}. Thus, for large $R$ and large $|A|$, $\xi(\hat{t})<|A|$ and critical property of $S$ for subsystem $A$ is dominated by the $R$, yielding Eqs.~(\ref{arealargeR}) and~(\ref{arealargeR1}), in which $|A|$ is absent.

\begin{figure}[tbp]
\centering
  \includegraphics[width=\linewidth,clip]{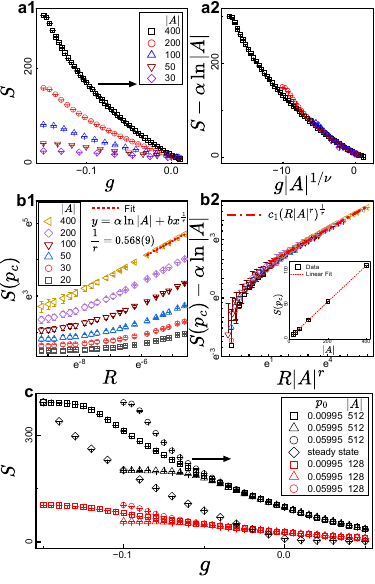}
  \vskip-3mm
  \caption{{\bf Driven critical dynamics of entanglement entropy $S$ from the volume-law phase.} (a) For a fixed $R|A|^r=9.292$, curves of $S$ for subsystem $A$ versus $g$ before (a1) and after (a2) rescaling. (b) Curves of $S$ versus $R$ at $p_c$ for different $A$ before (b1) and after (b2) rescaling. Double logarithmic scales are used. Fitting in (b1) for large $R$ shows $1/r=0.568(9)$, consistent with $1/r=0.557(3)$ with $\nu$ as input. Inset in (b2) shows that $S\propto |A|$ for $R=0.01$. Curve of $c_1(R|A|^r)^{1/r}$ (dash dot) is plotted for comparison in large $R|A|^r$ region in (b2). (c) Curves of $S$ for $|A|=512$ (black) and $|A|=128$ (red) versus $g$ with different initial states for $R=0.005$. Three different initial conditions are chosen: (i) the steady state at $p_0=0.00995$ (square), (ii) the steady state at $p_0=0.05995$ (triangle), and (iii) steady state of $p=0.00995$ but $p_0=0.05995$. The steady-state curve (diamond)  in the absence of driving for $|A|=512$ is also plotted for comparison. The arrows indicate the driving direction.
  }
  \label{fig:quench2}
\end{figure}

{\bf Driven dynamics from volume-law phase}--- Now we turn to the driven critical dynamics from the volume-law phase. The measurement probability is changed as $p=p_c+ R(t-t_c)$ with $p_0=0.00995$. At first, we examine Eq.~(\ref{genscaling}). Fig.~\ref{fig:quench2} (a) shows curves of $S$ versus $g$ for an arbitrary fixed $R|A|^r$. We find in Fig.~\ref{fig:quench2} (a2) that curves of $(S-\alpha{\rm ln}|A|)$ versus $g|A|^{1/\nu}$ collapse well onto each other, confirming that Eq.~(\ref{genscaling}) is applicable for the driven dynamics from both sides.

However, distinct asymptotic scaling behaviors appear when we focus on $S$ at $p_c$. In this case, Eq.~(\ref{genscaling}) also reduces to $S(p_c)=\alpha {\rm ln}|A|+G_1(R|A|^r)$, similar to the previous case, as confirmed by the scaling collapsing in Fig.~\ref{fig:quench2} (b2). But here for large $R$, $S(p_c)$ should be proportional to $|A|$ since the driving brings the information of the volume-law phase into the critical region, as confirmed in the inset of Fig.~\ref{fig:quench2} (b2). This property requires that $G_1$ must obey $G_1(R|A|^r)\sim (R|A|^r)^{1/r}$, as verified in Fig.~\ref{fig:quench2} (b2). For large $|A|$ and large $R$, $G_1(R|A|^r)$ can dominate over the logarithmic term. Accordingly, we obtain the scaling relation of $S(p_c)$ for large $R$,
\begin{equation}
S(p_c,R,|A|)\propto R^{1/r},
\label{volumelargeR}
\end{equation}
with the coefficient proportional to $|A|$.
By including an additional subleading logarithmic correction, power fitting in Fig.~\ref{fig:quench2} (b1) for the curve of $S$ versus large $R$ gives the exponent $0.568(9)$, which is close to the value of $1/r$ with $\nu$ and $z$ set as input, confirming Eq.~(\ref{volumelargeR}) and the scaling analyses. 


Besides, the dominant linear dependence of $S$ on $|A|$ in the scaling function can be generalized to the off-critical-point region by incorporating the steady case scaling, for which the scaling function obeys $F(g|A|^{1/\nu})\sim (|g||A|^{1/\nu})^{\nu}$ in the volume-law phase~\cite{LiydPhysRevB100134306}. Thus both scaling variables, $g|A|^{1/\nu}$ and $R|A|^r$, in $G$ dominate the physics for $g\leq 0$. A direct consequence is that although the variable $gR^{-1/\nu r}$ can emerge by the variable substitution formally, it cannot independently take on dominant physical effects in practice, in sharp contrast to the previous case of driving from the area-law phase. The fading of this $gR^{-1/\nu r}$ variable can undoubtedly result in the breakdown of the adiabatic-impulse scenario of the KZ mechanism, since it is just this term that takes responsibility for distinguishing the two scaling regions in usual phase transitions~\cite{Kibble1976,Zurek1985,delcamporev,Polkovnikov2011rmp,Dziarmaga2010review}. 

\begin{figure}[tbp]
\centering
  \includegraphics[width=\linewidth,clip]{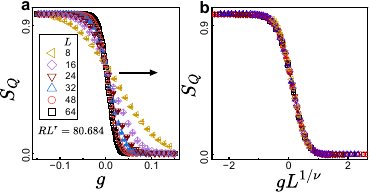}
  \vskip-3mm
  \caption{{\bf Driven critical dynamics of entanglement entropy $S_Q$ for auxiliary qubit from the volume-law phase.} For a fixed $RL^r=80.684$, curves of $S_Q$ for different system sizes $L$ versus $g$ before (a) and after (b) rescaling. The arrow shows the driving direction.
  }
  \label{fig:quench3}
\end{figure}

Furthermore, we find that the initial stage is neither an adiabatic region nor an impulse region. As $|A|$ is an extensive quantity, the indispensable term $R|A|^r$ makes the evolution of $S$ strongly depend on $R$ and deviate markedly from the steady state, even in the initial stage far from $p_c$, as shown in Fig.~\ref{fig:quench2} (c). Thus the evolution in the initial stage is nonadiabatic. Moreover, the initial stage is not an impulse region either, since the effects induced by different initial states can be eliminated in a transient time, as demonstrated in Fig.~\ref{fig:quench2} (c). Accordingly, this initial stage is a quasi-steady stage, which only depends on $g$, $|A|$ and $R$. More theoretical understanding on this quasi-steady stage can be found in the SM. The appearance of this $R$-dependent quasi-steady stage makes the driven dynamics of $S$ go beyond the conventional adiabatic-impulse scenario of KZ mechanism, but preserve the full FTS of Eq.~(\ref{genscaling}).



{\bf Discussion}--- Recently, the MIPT has been detected experimentally on quantum computers~\cite{Noelnpexp,Koh2023,Hoke2023}. In experiments, besides $S$, the entanglement entropy $S_Q$ of an ancillary qubit initially entangled to the system can also be used to diagnose the MIPT~\cite{Noelnpexp}. Here we investigate the driven dynamics of $S_Q$ from the volume-law phase for small sizes which are accessible in experiments. Note that the scaling dimension of $S_Q$ depends on the dynamic realization~\cite{HusePhysRevLett125070606,TurkeshiPhysRevB106214316}. Fig.~\ref{fig:quench3} shows that for fixed $RL^r$, $S_Q$ exhibits a drop near the MIPT. Moreover, we find $S_Q$ works as a dimensionless variable and its curves versus $g$ collapse onto each other after rescaling, demonstrating the FTS form for $S_Q$ is $S_Q=K(RL^r,gL^{1/\nu})$. It is quite promising that these results can be detected in experiments such as trapped-ion quantum simulators~\cite{Noelnpexp}.

Moreover, we note that the post-selection problem where the probability of obtaining identical trajectories vanishes exponentially in the number of measurements severely challenges the experimental investigation of emergent phenomena in monitored systems. 
Ref.~\cite{AltmanPRXQuantum5030311} has proposed to detect MIPT experimentally by approximating the entanglement entropy with appropriate cross-correlations which can be estimated with the help of classical simulations to mitigate the post-selection problem. However, the capacity of the classical computer still hinders the investigation of the volume-law regime. As shown in Eq.~\eqref{arealargeR1}, the area law dynamic scaling region can extend into the volume-law phase for $p<p_{c}$. Consequently, the driven critical dynamics can be naturally combined with the cross-correlation protocol to reveal the MIPT or other emergent phenomena in general monitored systems experimentally without the post-selection problem.


{\bf Summary}--- In this letter, we have explored the driven critical dynamics of the MIPT in a $(1+1)$D hybrid stabilizer circuits. For driven dynamics from the area-law phase, we have found the entanglement entropy $S$ for subsystem $A$ satisfies $S\propto \ln R$ at $p_c$ and is almost independent of the size of $A$ for large driving velocity. In this case, the dynamics obeys the adiabatic-impulse scenario of the KZ mechanism. In contrast, for driven dynamics from the volume-law phase, we have found that the adiabatic-impulse scenario breaks down and the initial stage now becomes an $R$-dependent quasi-steady state rather than the steady state in the absence of external driving. Moreover, at $p_c$, we have found a scaling relation $S\propto R^{1/r}$ for large $R$. Despite these differences, we have confirmed that the dynamics of the entanglement entropy $S$ can be described by a unified FTS form characterized by $R$. Moreover, we have also studied the scaling behaviors of $S_Q$. By generalizing the driven critical dynamics to the MIPT for the first time, our present work not only leads to a great leap to the fundamental theory of KZ mechanism and FTS, but also contributes to a new feasible approach to investigate the MIPT on realistic quantum devices~~\cite{Noelnpexp,Koh2023,Hoke2023}.

{\bf Acknowledgments}---W.W., J.L., and S.Y. are supported by the National Natural Science Foundation of China (Grants No. 12222515 and No. 12075324). S.Y. is also supported by the Science and Technology Projects in Guangdong Province (Grants No. 2021QN02X561). S.-X. Z. is supported by a startup grant at IOP-CAS.

\bibliographystyle{apsrev4-2}
\let\oldaddcontentsline\addcontentsline
\renewcommand{\addcontentsline}[3]{}
\bibliography{ref.bib}
\let\addcontentsline\oldaddcontentsline
\onecolumngrid

\clearpage
\newpage
\widetext

\begin{center}
\textbf{\large Supplemental Material for ``Driven Critical Dynamics in Measurement-induced Phase Transitions''}
\end{center}

\date{\today}
\maketitle

\renewcommand{\thefigure}{S\arabic{figure}}
\setcounter{figure}{0}
\renewcommand{\theequation}{S\arabic{equation}}
\setcounter{equation}{0}
\renewcommand{\thesection}{\Roman{section}}
\setcounter{section}{0}
\setcounter{secnumdepth}{4}

\addtocontents{toc}{\protect\setcounter{tocdepth}{0}}
{
\tableofcontents
}

\section{Finite size effects in driven dynamics in MIPT}
\subsection{General scaling form}
In the main text, we investigate the driven dynamics of $S$ for the subsystem $A$ embedded in a system with system size $L\rightarrow\infty$. Here we focus on the scaling effects induced by finite $L$. Scaling analysis combining finite-time scaling (FTS) and finite-size scaling (FSS) gives
\begin{equation}
S(R,g,|A|,L)=\alpha {\rm ln}|A|+G(R|A|^{r},g|A|^{1/\nu},L|A|^{-1}).
\label{genscalingfss}
\end{equation} 
When $R=0$, Eq.~(\ref{genscalingfss}) recovers the scaling form for MIPT in steady state~\cite{LiydPhysRevB100134306}. In particular, when $|A|=L/2$, $S$ corresponds to the half-chain entanglement entropy and its dynamic scaling reads
\begin{equation}
S(R,g,L)=\alpha {\rm ln}(L/2)+G_L(RL^{r},gL^{1/\nu}).
\label{genscalingfss1}
\end{equation}

In addition to $S$, the tripartite
quantum mutual information $I_3$, defined as
\begin{equation}
I_3\equiv S_A+S_B+S_C-S_{A\cup B}-S_{A\cup C}-S_{B\cup C}+S_{A\cup B\cup C},
\label{defmutuinf}
\end{equation}
in which ${A,B,C,D}$ is the quadripartition of the system and $|A|=|B|=|C|=|D|=L/4$, is also usualy used to characterize the MIPT~\cite{HusePhysRevB101060301}. It was shown that $I_3$ is a dimensionless variable at MIPT~\cite{HusePhysRevB101060301,TurkeshiPhysRevB106214316}. Accordingly, the FTS form of $I_3$ is
\begin{equation}
I_3(R,L,g)=Y(RL^{r},gL^{1/\nu}).
\label{genscalingfss2}
\end{equation}
In the following, we will explore the detailed properties included in these scaling forms.

\subsection{Driven critical dynamics from area-law phase}

\begin{figure}[tbp]
\centering
  \includegraphics[width=0.5\linewidth,height=8cm,clip]{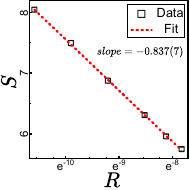}
  \vskip-3mm
  \caption{{\bf Driven dynamics of $S$ at $p_c$ from the area-law phase.} $S$ for half chain versus the driving velocity $R$. Semi-logarithmic scale is used. Linear fitting shows the slope is $-0.837$, which is close to the value of $\delta=-\alpha/r$ with $\alpha$, $\nu$ and $z$ set as input.
  }
  \label{fig:mipt_sm_area_figure1}
\end{figure}

At first, we investigate the driven critical dynamics of $S$ for half chain from the area-law phase. Here, $L$ is chosen as $L=1024$. We find that for large $R$, $S$ at $p_c$ satisfies
\begin{equation}
S(p_c)=\delta {\rm ln}R+c,
\label{genscalingfss3}
\end{equation}
in which $c$ is a constant. Note that this equation indicates that $S$ is almost independent of $L$ for large $R$. Fig.~\ref{fig:mipt_sm_area_figure1} shows the curve of $S$ versus $R$. In the semi-logarithmic scale, we find that the curve is almost a straight line, confirming $S\propto {\rm ln}R$. Moreover, linear fitting shows that the slope is close to the value of $\delta$, confirming Eq.~(\ref{genscalingfss3}). 

Moreover, at $p_c$, Eq.~(\ref{genscalingfss1}) reduces to
\begin{equation}
S(R,g,L)=\alpha {\rm ln}(L/2)+G_{L1}(RL^{r}).
\label{genscalingfss4}
\end{equation}
Eq.~(\ref{genscalingfss3}) dictates that $G_{L1}(RL^{r})$ obeys $G_{L1}(RL^{r})=\delta {\rm ln}(RL^{r})+c$. The physical reason is that for driving from the area-law phase, when $R^{-1/r}<L$, the correlation length is smaller than $L$. Thus, the frozen length scale governs the scaling of $S$. These results are consistent with those for large $A$ and large $R$ discussed in the main text.

\begin{figure}[tbp]
\centering
  \includegraphics[width=0.9\linewidth,clip]{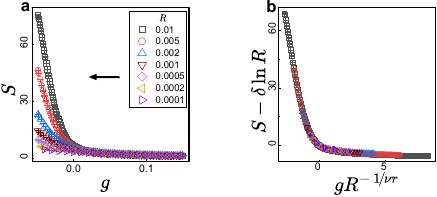}
  \vskip-3mm
  \caption{{\bf Driven dynamics of $S$ from the area-law phase.} $S$ for half chain versus $g$ for different $R$ before (a) and after (b) rescaling. The arrow indicates the driving direction.
  }
  \label{fig:mipt_sm_area_figure2}
\end{figure}

To generalize the scaling relation for $g\neq 0$, we explore the dynamic scaling of $S$ in the driven process. For large $R$, the scaling form of $S$ should be independent of $L$. By incorporating Eq.~(\ref{genscalingfss3}), Eq.~(\ref{genscalingfss2}) can change to
\begin{equation}
S(R,g,L)=\delta {\rm ln}(R)+G_{L2}(gR^{-1/\nu r}).
\label{genscalingfss5}
\end{equation}
Fig.~\ref{fig:mipt_sm_area_figure2} shows that the evolution of $S$ versus $g$ for different driving rate $R$. After rescaling the variables according to Eq.~(\ref{genscalingfss5}), the rescaled curves collapse well, verifying Eq.~(\ref{genscalingfss5}).

\begin{figure}[tbp]
\centering
  \includegraphics[width=0.9\linewidth,clip]{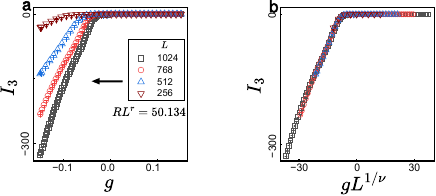}
  \vskip-3mm
  \caption{{\bf Driven dynamics of $I_3$ from the area-law phase.} $I_3$ versus $g$ for  different $L$ with a fixed $RL^{r}=50.134$ before (a) and after (b) rescaling. The arrow indicates the driving direction.
  }
  \label{fig:mipt_sm_area_figure3}
\end{figure}

Then we investigate the dynamic scaling of $I_3$. We find that $I_3$ is more dependent on $L$ than $S$. A possible reason is that the relevant linear scale in $I_3$ is a quarter of $L$, which is shorter than $L/2$ for $S$. Accordingly, we resort to the FTS form with the finite-size effects included. Fig.~\ref{fig:mipt_sm_area_figure3} shows that for an arbitrarily fixed $RL^{r}=50.134$, curves of $I_3$ versus $g$ for different $L$ collapse onto each other when $g$ is rescaled as $gL^{1/\nu}$, confirming Eq.~(\ref{genscalingfss2}).

By inspecting the curves of $S$ shown in Fig.~\ref{fig:mipt_sm_area_figure2} and the curves of $I_3$ shown in Fig.~\ref{fig:mipt_sm_area_figure3}, we find that in the initial stage of the evolution, these curves are all independent of $R$, demonstrating the evolution is in the adiabatic stage. Then they deviates from each other when $p$ gets close to the $p_c$, demonstrating the evolution is in the impulse stage. Therefore, the driven dynamics from the area-law phase satisfies the adiabatic-impulse scenario of the original KZ mechanism. 

\begin{figure}[tbp]
\centering
  \includegraphics[width=0.9\linewidth,clip]{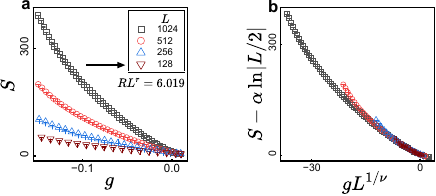}
  \vskip-3mm
  \caption{{\bf Driven dynamics of $S$ from the volume-law phase.} $S$ versus $g$ for  different $L$ with a fixed $RL^{r}=6.019$ before (a) and after (b) rescaling. The arrow indicates the driving direction.
  }
  \label{fig:mipt_sm_volume_figure1}
\end{figure}

\subsection{Driven dynamics from volume-law phase}
In this section, we study the driven dynamics from the volume-law phase. At first, we examine the full scaling form of $S$ for half system, i.e., Eq.~(\ref{genscalingfss1}). For an arbitrarily fixed $RL^{r}=6.019$, we find that curves of $S$ versus $g$ match with each other when $S$ and $g$ are rescaled according to $(S-\alpha {\rm ln}(L/2))$ and $gL^{1/\nu}$, respectively, confirming Eq.~(\ref{genscalingfss1}).

\begin{figure}[tbp]
\centering
  \includegraphics[width=0.9\linewidth,clip]{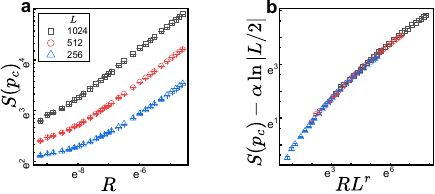}
  \vskip-3mm
  \caption{{\bf Driven dynamics of $S$ at $p_c$ from the volume-law phase.} $S$ versus $R$ for  different $L$ before (a) and after (b) rescaling. Double logarithmic scales are used.
  }
  \label{fig:mipt_sm_volume_figure2}
\end{figure}

Then we explore the scaling property of $S$ at $p_c$. In this case, Eq.~(\ref{genscalingfss1}) reduces to 
\begin{equation}
S(R,g,L)=\alpha {\rm ln}(L/2)+G_{L3}(RL^{r}).
\label{genscalingfss6}
\end{equation}
For different $L$, Fig.~\ref{fig:mipt_sm_volume_figure2} shows that rescaled curves of $S(p_c)$ versus $R$ collapse together according to Eq.~(\ref{genscalingfss6}), confirming that the dependence of $S(p_c)$ on $R$ for different size satisfies Eq.~(\ref{genscalingfss6}).

\begin{figure}[tbp]
\centering
  \includegraphics[width=0.5\linewidth,height=8cm,clip]{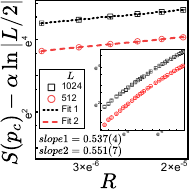}
  \vskip-3mm
  \caption{{\bf Driven dynamics of $S$ at $p_c$ from the volume-law phase.} Curves of $S-\alpha {\rm ln}(L/2)$ versus $R$ for two different $L$ are plotted in double logarithmic scales. Power fitting shows that the exponents (slope in double logarithmic scales) are both close to $1/r$ with $\nu$ and $z$ set as input. Inset shows broader $R$-region for both sizes.
  }
  \label{fig:mipt_sm_volume_figure3}
\end{figure}

Eq.~(\ref{genscalingfss6}) is similar to Eq.~(\ref{genscalingfss4}) in form. But we will find that the asymptotic behaviors of their scaling functions are very different. Fig.~\ref{fig:mipt_sm_volume_figure3} shows that the curve of $S-\alpha {\rm ln}(L/2)$ versus large $R$ in double logarithmic scales is almost a straight line, indicating that $S-\alpha {\rm ln}(L/2)$ should be a power function of $R$. Power fitting shows that the exponents for different $L$ are both close to $1/r$. These results demonstrates that $G_{L3}(RL^r)$ satisfies $G_{L3}(RL^r)\sim (RL^r)^{1/r}$. Accordingly, $G_{L3}(RL^r)$ strongly depends on $L$ for large $R$. Thus, $G_{L3}(RL^r)$ is remarkably different from $G_{L2}(RL^r)$, which is almost independent of $L$.

\begin{figure}[tbp]
\centering
  \includegraphics[width=0.9\linewidth,clip]{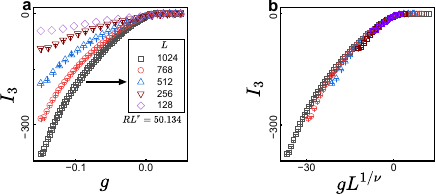}
  \vskip-3mm
  \caption{{\bf Driven dynamics of $I_3$ from the volume-law phase.} $I_3$ versus $g$ for  different $L$ with a fixed $RL^{r}=50.134$ before (a) and after (b) rescaling. The arrow indicates the driving direction.
  }
  \label{fig:mipt_sm_volume_figure4}
\end{figure}

We also investigate the dynamic scaling of $I_3$ from the volume-law phase. Fig.~\ref{fig:mipt_sm_volume_figure4} shows that for an arbitrarily fixed $RL^{r}=50.134$, curves of $I_3$ versus $g$ for different $L$ collapse onto each other when $g$ is rescaled as $gL^{1/\nu}$, confirming Eq.~(\ref{genscalingfss2}).

Moreover, by comparing Figs.~\ref{fig:mipt_sm_volume_figure1} and \ref{fig:mipt_sm_volume_figure4} with those from the area-law cases, we find that for the driven dynamics from the volume-law phase, the initial stage is not an adiabatic stage. Thus, the original adiabtic-impulse scenario of the KZ mechanism breaks down, although the FTS forms are still applicable. 

\section{Theoretical understanding of the quasi-steady stage}
As discussed in the main text, there is a quasi-steady stage in the driven dynamics starting from the volume law phase. In this section, we show how to understand this quasi-steady stage in terms of the effective statistical model.

\begin{figure}[tbp]
\centering
  \includegraphics[width=0.9\linewidth,clip]{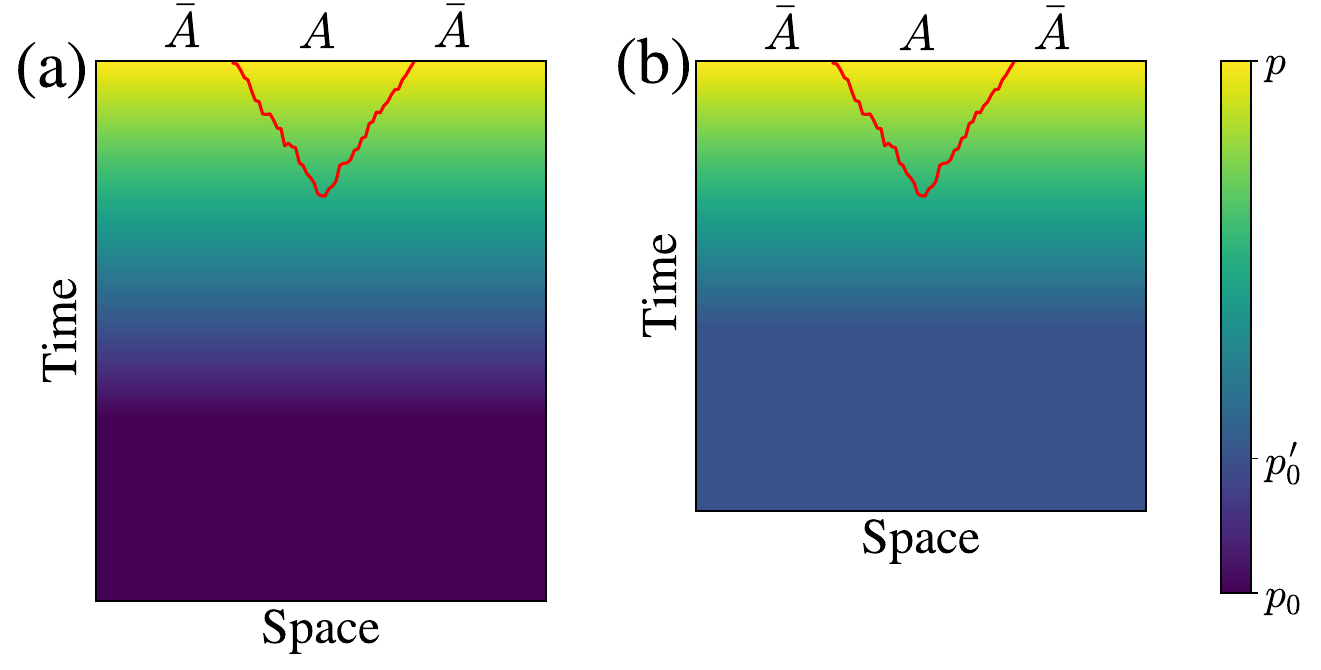}
  \vskip-3mm
  \caption{{\bf Theoretical understanding of the quasi-steady stage.} (a) and (b) show the corresponding statistical model of the driven critical dynamics with initial probability $p_{0}$ and $p_{0}^{\prime}$ respectively. Color represents the measurement probability, i.e., the strength of the attractive potential. When $t$ is large, the initial uniform potential is far away from the top boundary and thus the domain wall configuration depends on the final probability $p$ and driving velocity $R$. There is a quasi-steady stage independent on the initial measurement probability $p_{0}$.
  }
  \label{fig:quasi_steady_stage}
\end{figure}

It is well known that the $(1+1)$D random unitary circuits can be mapped to a 2D classical model with ferromagnetic spin-spin interaction~\cite{PhysRevB.99.174205,BaoPhysRevB101104301,JiancmPhysRevB101104302, PhysRevLett.129.080501,LiusPhysRevLett132240402}. The entanglement entropy of subsystem $A$ in the steady state of the random unitary circuits corresponds to the free energy of the classical spin model with specific top boundary conditions, which is proportional to the length of the unique domain wall. The projective measurements with random space-time locations act as a random attractive potential with a strength determined by the measurement probability $p$ in the classical spin model,
and thus the presence of projective measurements will induce the fluctuation of the domain wall. With a constant measurement probability, the strength of the attractive potential is uniform, while if the measurement probability is time-dependent, the strength of the attractive potential depends on the vertical distance away from the top boundary as shown in Fig.~\ref{fig:quasi_steady_stage}. For the driven dynamics starting from the steady state with initial measurement probability $p_{0}$, in the early time, the uniform attractive potential dominates, and thus the domain wall configuration, as well as the entanglement entropy, depends on the initial $p_{0}$. However, as $t$ increases, the uniform potential is far away from the top boundary and thus has no influence on the domain wall configuration as illustrated in Fig.~\ref{fig:quasi_steady_stage}. Therefore, the domain wall configuration as well as the entanglement entropy is determined by the time-dependent potential which only depends on the final measurement probability and the driving velocity $R$. Consequently, there is an $R$-dependent quasi-steady stage.

\end{document}